\documentclass{elsart}

\usepackage{natbib}

\usepackage{graphicx}
 
\begin{document}


\runauthor{Garrington, Garrett  and Polatidis}


\begin{frontmatter} 
\title{A VLBI and MERLIN Survey of faint, compact radio sources}

\author[NRAL]{S.T. Garrington}
\author[JIVE]{M.A. Garrett}
\author[JIVE,OSO]{A. Polatidis}

\address[NRAL]{Nuffield Radio Astronomy Laboratories, The University of Manchester, Jodrell Bank, Macclesfield SK11 9DL, UK}
\address[JIVE]{Joint Institute for VLBI in Europe, Postbus 2, 
7990~AA Dwingeloo, The Netherlands} 
\address[OSO]{Onsala Space Observatory, Onsala, Sweden}
 

\begin{abstract} 
  
  We present preliminary results from a VLBI survey at
  \mbox{$\lambda=6\,$cm} of a sample of 35 sources with flux densities
  of \mbox{2 -- 100 mJy}. These sources were selected from the VLA
  FIRST survey at \mbox{$\lambda = 20\,$cm}, in a 3 degree field around
  the bright calibrator 1156+295, simply by imposing \mbox{$S_{20} > 10
  $ mJy} and \mbox{$\theta < 5$ arcsec.} MERLIN observations at
  \mbox{$\lambda 6\,$cm} detected 70/127 of these sources with a
  threshold of 2 mJy at 50 mas resolution and the closest 35 of these
  to the calibrator were observed with the VLBA+EVN in snapshot mode at
  $\lambda 6\,$cm.  These sources are a mixture of flat and
  steep-spectrum sources and include: weak flat-spectrum nuclei of
  large radio galaxies, low power AGN in nearby galaxies and radio
  quiet quasars.  With these short observations, the sensitivity is
  limited and most appear as either core-jets or simple point sources
  on the milliarcsec scale.  Nonetheless, it is encouraging that with
  only 10 minutes observation per source, at least 35\% of {\em all}
  sources with \mbox{$S_{20} > 10 $ mJy} can be detected and imaged
  with global 6cm VLBI.

\end{abstract} 



\end{frontmatter} 

\section{Introduction}
\label{intro} 

The new VLA surveys, FIRST \citep{first}  and NVSS \citep{nvss}, 
have opened up the mJy radio population on a vast scale, but even the
higher resolution FIRST survey does not resolve the majority of
sources which it detects.  The tendency for the fainter sources to be
considerably smaller has been known for some years \citep{oort} and
is strikingly demonstrated by recently compiled statistics of the
large MIT VLA surveys, where 47\% of the sources brighter than 50 mJy 
imaged with 0.25 arcsec resolution at 8.4 GHz were found to be unresolved
\citep{mit}

However, this mJy population has yet to be systematically investigated
with VLBI, previous VLBI surveys have so far been limited to flux
densities $ > $ 100 mJy \citep{cj,vlbacal} and often these samples have
been restricted to relatively flat spectrum sources ($\alpha < 0.5$).
High resolution images will be of interest for population studies of
weak AGN, such as the evolutionary schemes recently proposed by
\citet{wall}, and will be a valuable input to statistical studies of
gravitational lensing, since the compact mJy radio source population is
the `parent population' of the lensed sources being detected in the
CLASS and MIT surveys \citep{kochanek}.

\begin{figure}[hbt] 
\centering
\includegraphics[scale=0.4,angle=0]{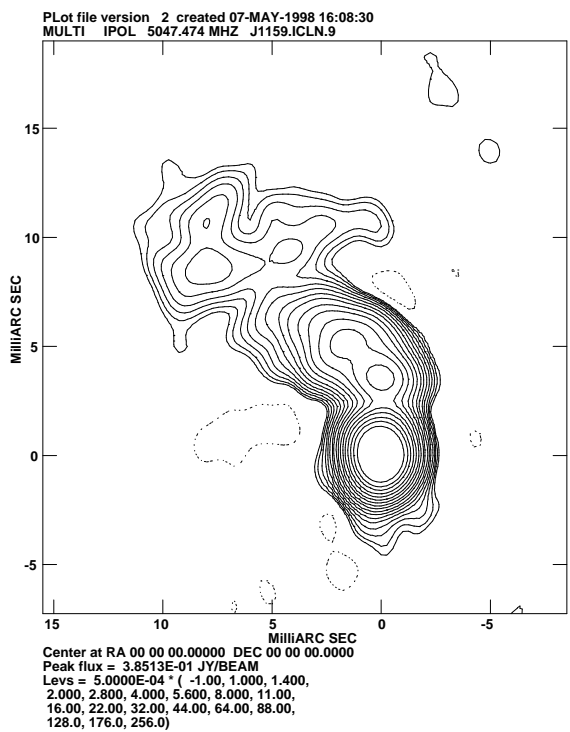}  
\caption{Global 6cm VLBI image of 1156+295, used as the phase calibrator.}
\label{fig1} 
\end{figure}  

VLBI observations of mJy sources are only possible by using the
technique of phase referencing. The density of the new VLA surveys
is such that we are now free to choose an ideal phase calibrator
source and then simply select targets from the surrounding field.

The initial swathe of the FIRST survey included the area around
1156+295, an interesting source in its own right (Figure 1, see also
Hong, these proceedings) but also an ideal phase calibrator. We
therefore defined a $3 \times 4$ degree field centred on this source
and simply selected as potential targets all 127 sources from the FIRST
catalogue in this area with $S_{1.4} > 10$ mJy and LAS $<$ 5.0 arcsec.

\section{MERLIN Observations }

In order to select those sources with sufficient emission on
milliarcsec scales for VLBI imaging, we used MERLIN at $\lambda
6\,$cm, observing each source for $3 \times 5\,$ minutes, giving a
detection threshold of about 2 mJy with 50 mas resolution \citep{G98}.
This is a savage cut for steep-spectrum sources, which would have to
be very compact to be detected by MERLIN at this level, but we would
expect to detect the majority of flat-spectrum sources, which should
account for $ < 30$\% of the sample.  Perhaps surprisingly, just over
half of the list, 70 sources, were detected.  These appeared to be
mostly point-sources, but there were a few simple double sources, and
several marginal detections of complex emission from the lobes of
large double sources, which, given the very simple selection criteria,
are not excluded from the target list.

VLA A-configuration observations at 1.4 and 8.4 GHz (1.0 and 0.25
arcsec resolutions) were made of the final target list in order
to image  structure on intermediate scales and to determine spectral
indices. MERLIN + EVN observations at 18cm have also been made and and
the EVN data are now being correlated.

\section{Global VLBI Observations at $\lambda = 6\,$cm}

Global $\lambda 6\,$cm VLBI observations (18 telescopes; VLBA + EVN)
were made earlier this year. To meet scheduling constraints, 35 sources
closest to the phase calibrator and detected by MERLIN at 6cm were
selected.  With $6 \times 2$ minutes observation per source, we
achieved 2 mas resolution and a detection threshold of 1 - 2~mJy. The
source positions were known to within a few mas from the MERLIN
observations, reducing the risk of false detections.

Only 8 of the 35 sources were not detected in these global VLBI
observations. These were bright, steep-spectrum and clearly extended
on intermediate scales  three were in fact lobes of large double sources,
four were CSS sources with sizes $>$ 200 mas and one source was a
marginal detection with MERLIN. Indeed, we would not have expected to
have detected these sources, but they were not excluded by the 'blind'
selection process.

\begin{figure}[hbt] 
\centering
\includegraphics[scale=1.2,angle=0]{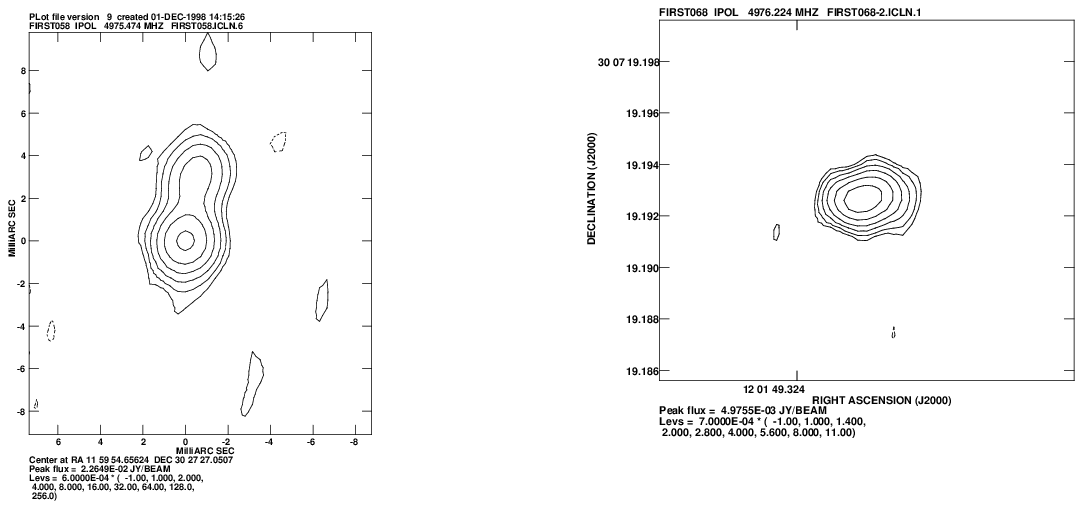}
  
\caption{These faint milliarcsec flat spectrum sources are 
associated with bright (R=14 and 16) galaxies.  Neither have significant 
extended radio
emission on scales larger than this.}
\label{fig2} 
\end{figure}  

Of the 27 sources which were detected, there were roughly equal
numbers of flat and steep-spectrum sources. 

The brighter flat-spectrum sources are associated with nearby
red galaxies. Two  are the nuclei of arcminute scale  radio galaxies,
but otherwise they have no extended emission on intermediate scales
(0.1 - 10 arcsec), and two are quasars. 
On the milliarcsecond scale, these sources
mostly have core-jet structures.  The fainter sources (10 - 50 mJy)
have no optical identifications on the POSS and appear to be
unresolved from mas to arcsecond scales.  These flat-spectrum
sources have $T_{\rm b} \sim 10^{10} - 10^{11}\,$ K.

\begin{figure}[hbt] 
\centering
\includegraphics[scale=1.2,angle=0]{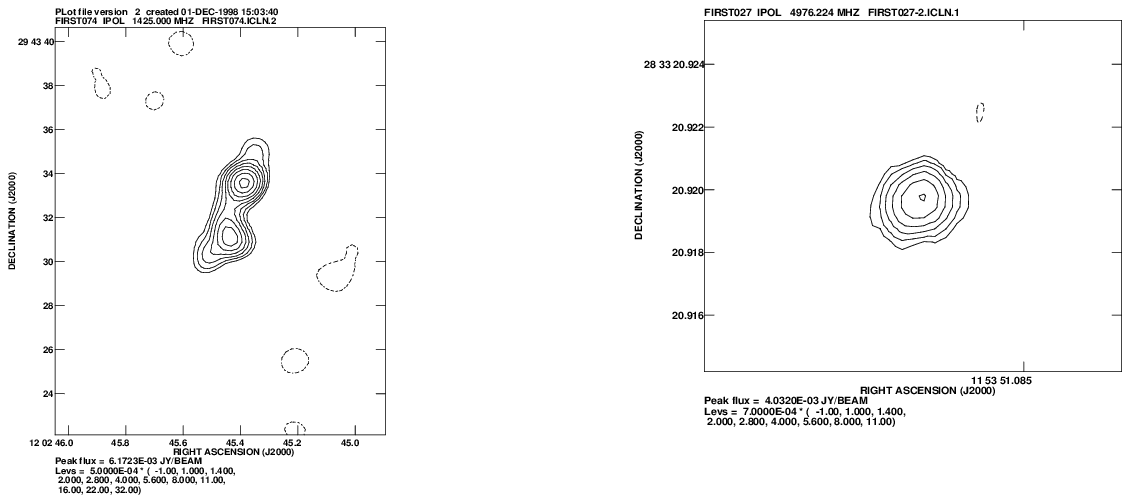}
  
\caption{The VLA A-configuration 20cm map on the left shows a flat spectrum
core and an extended 5 arcsec jet to the south in this 18 mJy source.
There may also be lower surface brightness features resolved out
by this image with a total flux density of a few mJy.  The 6cm global VLBI
phase referenced image on the right, shows the 5 mJy core which is barely
resolved.  The slight extension may be due to residual phase errors
and a realistic upper limit on the component size is 1 mas, confirmed using
the VLBA data alone. This source may be identified with one
of a group of faint galaxies.}

\label{fig3} 
\end{figure}  

Of the steep-spectrum sources, six are barely resolved: these
may be  faint GPS sources with spectral peaks below 1 GHz
and expected sizes of 1 - 10 mas (Snellen, these proceedings).
If so, this would indicate quite a high fraction of GPS sources at
this flux level.  The sources with intermediate spectral indices
have, as expected, arcsecond scale jets and mas cores. The largest
example is a 5 arcsecond triple source, which just fell within the 
initial selection.

\begin{figure}[hbt] 
\centering
\includegraphics[scale=0.4,angle=0]{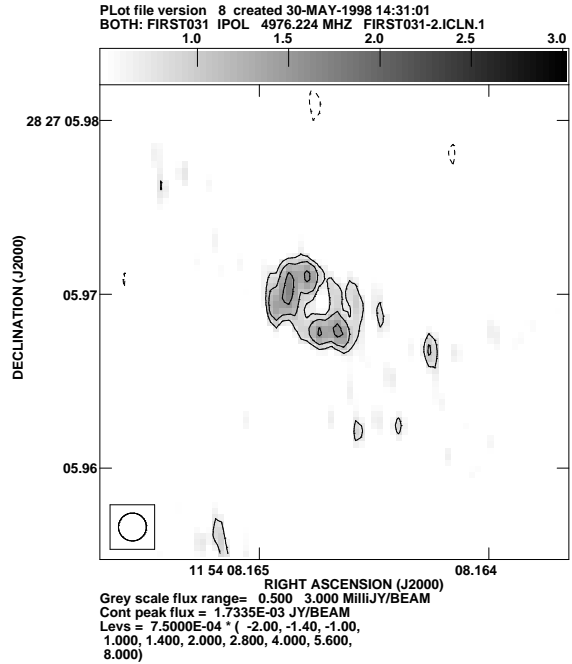}
  
\caption{
  This steep spectrum source has no optical identification on the POSS
  plate and no significant extended emission on arcsec scales. The
  phase-referenced image shows a complex, ring-like structure.  }

\label{fig4} 
\end{figure}

One source has a rather unusual 5 mas diameter ring-like structure.
Optically, this is an empty field down to 21 mag.  The nearest optical
objects are 30 arcsec away and are faint. The source  has a spectral
index of 0.80 with no extended radio structure on any scales from 50
mas to several arcminutes, except perhaps a few mJy within 0.5 arcsec
of the `core' and its nearest radio neighbour is a 4 mJy source 2
arcminutes away.  Ascribing this distorted structure to gravitational
lensing would imply a lens mass of $\sim 10^6$ solar masses. The expected
probability for lensing on this scale, even assuming all normal
galaxies harbour massive black holes is tiny ($3 \times 10^{-9}$).
Alternatively, this may be an extremely distorted jet, perhaps like
3C119 \citep{3c119} or Mkn 501 \citep{mrk501} but 50 times fainter and
several times smaller. More sensitive VLBI observations are required to
investigate this source further.

\section{Conclusions}

We have produced VLBI images of an un-biased sample (no spectral index
or optical selection) of faint, compact radio sources.  With only 10
minutes observation per source, approximately 35\% of {\em all} sources
with $S_{20} > 10$  mJy can be detected on  global VLBI baselines at
6cm.

This project has used the simplest of selection criteria and we have
seen that that at the mJy level, the VLBI sky shows a broad range of
flat and steep spectrum sources, apparently distant and nearby.  In
order to complete the statistical picture on this small sample, we are
awaiting optical imaging and spectroscopic observations as well as the
EVN 18cm observations to complete radio picture.  Perhaps the most
important lesson from this pilot VLBI survey of faint radio sources is
that the mJy population is readily accessible with current VLBI
techniques.  The new VLA surveys provide a huge resource for this type
of work and MERLIN can be used as a very efficient filter for selecting
potential VLBI targets.

\end{document}